# Surface structuring of patterned 4H-SiC surfaces using a SiC/Si/SiC sandwich approach

Y. Jousseaume[1a], P. Kumar[2b], M.E. Bathen[2c], F. Cauwet[1d], U. Grossner[2e], G. Ferro[1f]

[1]Laboratoire des Multimatériaux et Interfaces, Université de Lyon, 6 rue Victor Grignard, 69622 Villeurbanne, France
[2] Advanced Power Semiconductor Laboratory, ETH Zurich, Physikstrasse 3, 8092 Zurich, Switzerland

[a]yann.jousseaume@univ-lyon1.fr, [b]kumar@aps.ee.ethz.ch, [c]marianne.bathen@gmail.com, [d]francois.cauwet@univ-lyon1.fr, [e]ulrike.grossner@ethz.ch, [f]gabriel.ferro@univ-lyon1.fr

**Abstract.** Mesa- and trench-patterned surfaces of 4H-SiC(0001) 4°off wafers were structured in macrosteps using Si melting in a SiC-Si-SiC sandwich configuration. Si spreading difficulties were observed in the case of trench-patterned samples while the attempts on mesa-patterned ones were more successful. In the latter case, parallel macrosteps were formed on both the dry-etched and un-etched areas though these macrosteps rarely cross the patterns edges. The proposed mechanism involved preferential etching at Si-C bilayer step edges and fast lateral propagation along the $[11\bar{2}0]$ direction.

**Introduction**

Recently, there have been several interests in studying the surface structuration of 4H-SiC(0001) 4°off wafers into macrosteps, either for reducing electrically active defects at the $SiO_2$/4H-SiC interface in metal-oxide-semiconductor field effect transistors (MOSFET) [1-3], for spatially localizing single photons emitters at the step edges [4] or for understanding the behavior of macrosteps towards liquid phase growth optimization [5]. Using a SiC/Si/SiC sandwich approach, it was shown that regular and parallel macrosteps (of 3-5 μm width) could be obtained on 4H-SiC wafers [3]. The proposed mechanism at the origin of this surface structuration involves dissolution of the bottom SiC wafer, due to the vertical thermal gradient inside the stack. This dissolution naturally generates a step bunching whose parallelism can be reproducibly controlled via the prior deposition of an epitaxial layer thanks to the pre-structuration in parallel microsteps of this kind of surface. In the present work, we investigate the use of patterned 4H-SiC surfaces in order to check if this sandwich approach can be applied to an already processed surface and/or to see how it affects the structuring mechanism. The goal is to better understand this structuring mechanism while determining the limits of the sandwich approach.

**Experimental**

The SiC/Si/SiC sandwich consists of a bulk piece of silicon wafer which is melted between two 4H-SiC (0001) 4° off Si-face wafers (Fig. 1). The top SiC wafer has the role of homogeneously spreading the liquid Si by pressing and wetting. The amount of Si was chosen so as to lead to a 30 μm thick liquid film after melting. The bottom SiC wafer was chosen larger than the top one (i.e. 2x2 $cm^2$ and 1.2x1.2 $cm^2$, respectively) to avoid any Si loss from the SiC wafer sides upon melting. The heat treatments were performed in a homemade vertical cold wall reactor working at atmospheric pressure under 12 slm of $H_2$. The stacks were placed on a SiC-coated graphite susceptor and RF-heated, with a ramp up rate of ~400°C/min up to 1550°C. This temperature was kept for 30 min before fast natural cooling under $H_2$. After the treatments, the samples were wet etched in concentrated $HF/HNO_3$ for removing the Si and thus separating both wafers.

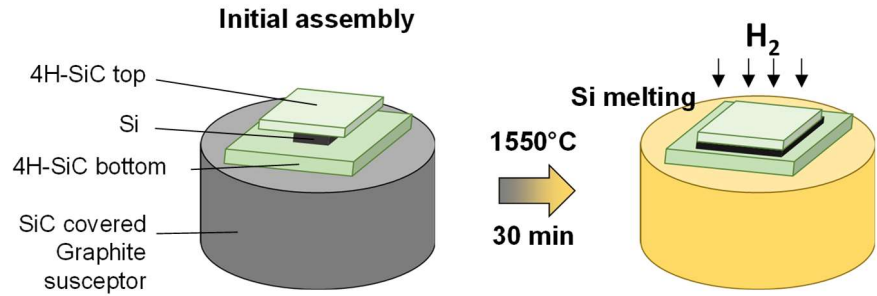

**Figure 1.** Schematic representation of the SiC/Si/SiC assembly before and after Si melting.

In order to check the effect of surface patterning on the structuring process, only the surface of the bottom SiC wafer was processed. A $n^+$ doped 4H-SiC wafer having received a 10 μm thick homoepitaxial layer with n- doping $< 1 \times 10^{17}$ cm$^{-3}$ was patterned using standard photolithography. Both positive and negative resins were tested for fabricating respectively rectangular trenches (2.5 μm deep, sample A) and rectangular mesas (1.5 μm high, sample B). Note that several samples of both A and B kinds were made for testing the reproducibility of the results. The morphology of the un-etched areas (epilayer surface) and the dry-etched areas (between the mesas or inside the trenches) are shown in Fig. 2. One can see that they are very different: the epilayer surface is composed of elongated microsteps while the etched areas display circular prints of several nm depth.

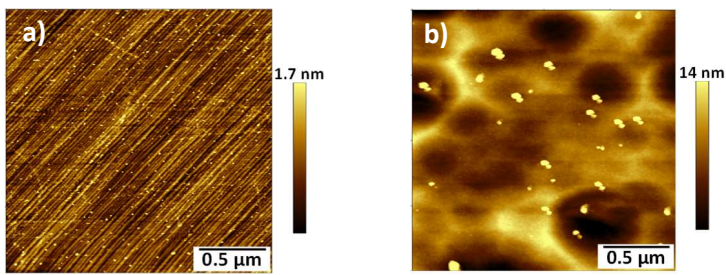

**Figure 2.** AFM images (2.5x2.5 μm$^2$) of typical a) un-etched area and b) dry-etched area of both samples A and B before surface structuration experiments.

**Results and discussion**

Let us start with sample A (trench pattern). The results obtained after the structuration treatment on such a sample are shown in Fig. 3. One can see that the liquid Si did not spread properly and did not wet all the area occupied by the SiC top wafer (Fig. 3a, b). Many circular features of few hundreds of μm diameter can be seen on the surface of the bottom SiC wafer (Fig. 3c). They are probably the result of gas trapping, a phenomenon which has not previously been observed for standard (non-patterned) surfaces treated with the same conditions [3,6]. All these observations suggest some difficulties in the wetting and spreading of the liquid Si when using trench patterns. This was repeatedly found on all the samples of this kind.

Beside this, elongated and parallel macrosteps are formed but they do not cover homogeneously the surface. Surprisingly, these parallel macrosteps appear on both dry-etched (trench) and un-etched areas despite the difference of initial surface morphology of these areas (Fig. 2). One can see that the macrosteps outside the trenches rarely propagate through the trenches. They seem to be blocked at the pattern's edges (this is better seen in Fig. 3d). Note that some macrosteps can be also seen inside the circles meaning that the liquid was also locally in contact with the SiC wafer after a certain time. The associated gas trapping should be thus a transient phenomenon.

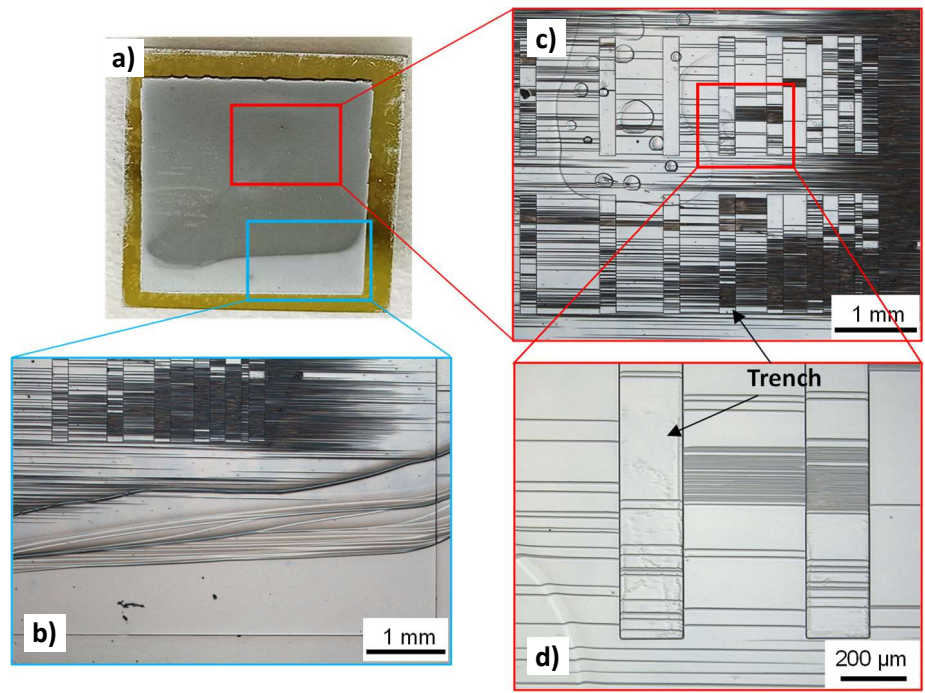

**Figure 3.** Characterization of sample A after surface structuration treatment: a) picture of the complete sample before Si etching and b-d) optical microscopy images of the bottom SiC wafer after etching; b) zoom-in on the blue area in a), c) zoom-in on the red area in a) and d) zoom-in on the red area in c).

Coming now to sample B, the results obtained after its structuration treatment are shown in Fig. 4. No specific problem of wetting can be seen, either macroscopic or microscopic (circular features). This result was reproducibly obtained with all the samples of this kind. The structuration process with mesa patterns behaves thus similarly as with the un-patterned surfaces. Besides this, sample B looks rather the same as sample A, with parallel and elongated steps forming on both dry etched and un-etched areas. Again, the macrosteps outside the patterns rarely propagate through the patterns. This is better seen in Fig. 5 showing the case of a mesa-patterned sample with mesas mostly oriented perpendicular to the macrostep edges. One can clearly see that the mesas are obstacles to the lateral propagation of these steps. As a consequence, the density of macrosteps in the dry-etched areas looks higher in the zones where mesas do not make obstacles. Furthermore, both the width and the in-plane orientation of the rectangular mesas seem to have an effect on the macrosteps propagation, but one would need more statistics to draw any conclusion. When looking closer to the morphology of the non-structured areas (either on or outside mesas), one can see undulated features forming perpendicularly to the step front (AFM of Fig. 5). This morphology was never observed on similarly treated un-patterned surfaces.

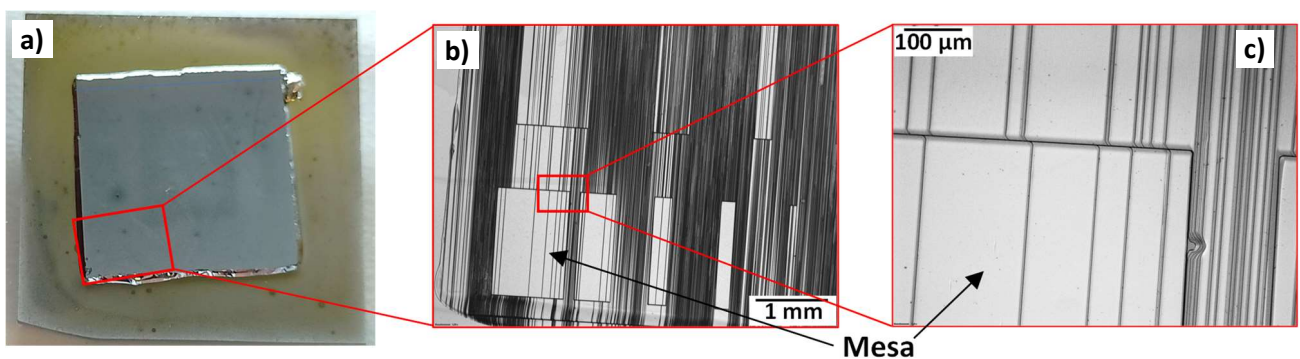

**Figure 4.** Characterization of sample B after surface structuration treatment: a) picture of the complete sample before Si etching and b-c) optical microscopy images of the bottom SiC wafer after etching; b) zoom-in on the red area in a), c) zoom-in on the red area in b).

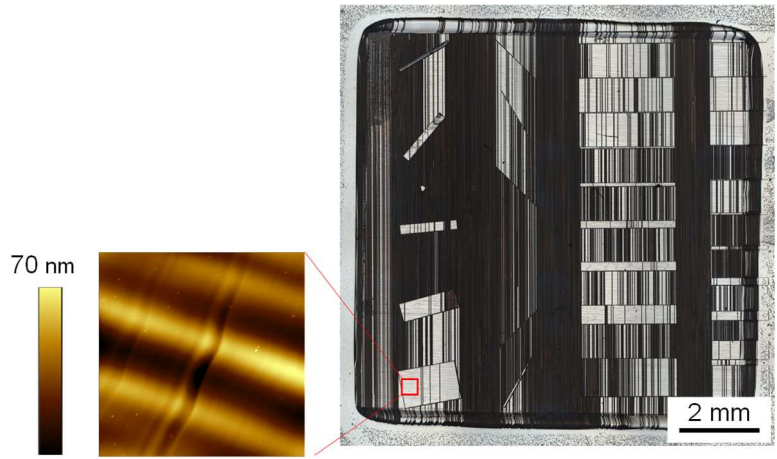

**Figure 5.** Image of the entire reconstructed area of a B type sample with mesas mostly oriented perpendicularly to the macrosteps edges. This image was obtained by multiple images alignment (MIA) of optical microscope images. The 50x50 μm² AFM scan shown on the left was recorded on the red square area of the MIA image.

Mechanical profilometry mappings were performed on the bottom SiC wafer of sample A (same as shown in Fig. 4), before and after the thermal treatment (Fig. 6a, c). The mesas can be easily distinguished in both cases, their average vertical and lateral dimensions being very similar (Fig. 6b, d). During the structuration process, the bottom SiC wafer undergoes dissolution (estimated to be~ 0.75 μm after 30 min, as deduced from other experiments [3]). Obviously, this dissolution occurs identically on the entire surface (on or outside the mesas) without any visible anisotropy. Longer experiments would probably allow refining these observations. Interestingly, the surface roughness outside the mesas has significantly decreased after structuration. As a matter of fact, the process leading to macrosteps formation has some long-range smoothening effect on the dry-etched areas.

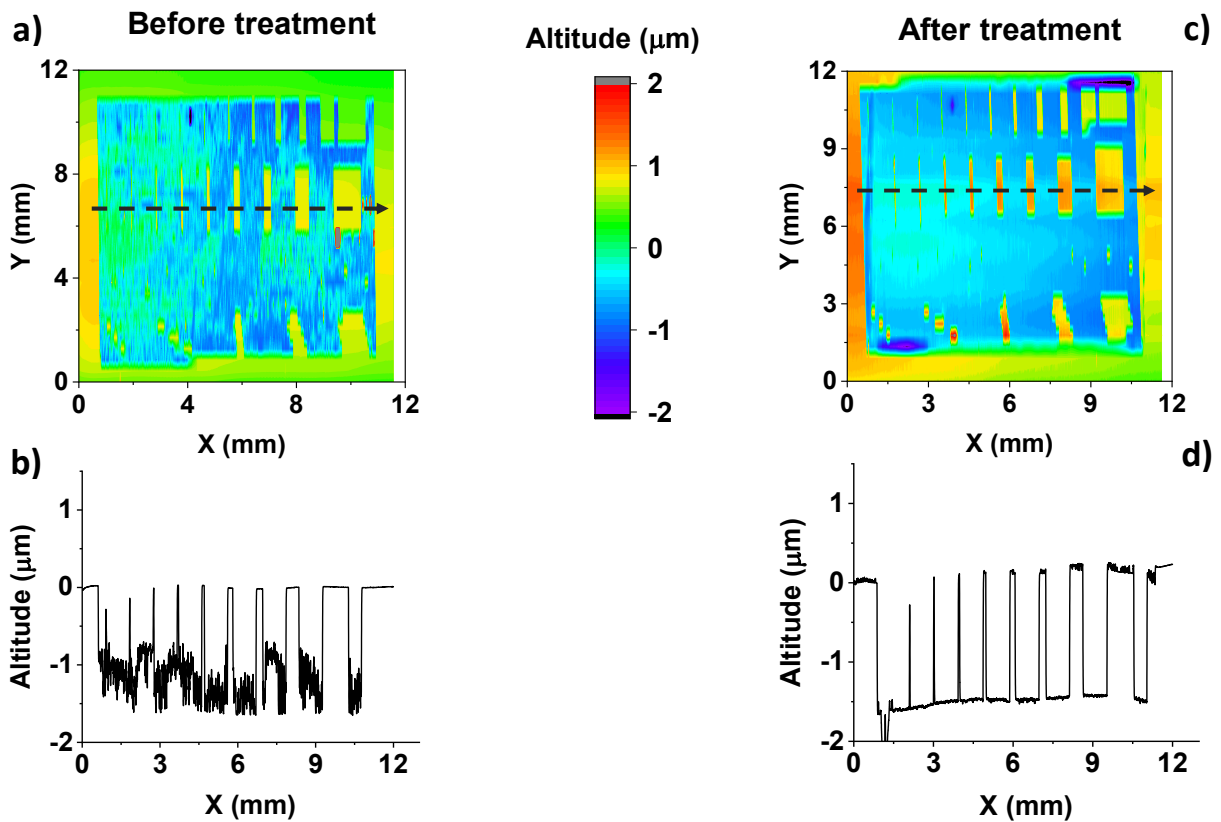

**Figure 6.** Mechanical profiling of the entire reconstructed area of sample B a) before and c) after thermal treatment.; c) linear profile along the black arrow in a), and d) linear profile along the black arrow in c).

From these observations, it is obvious that the mechanism leading to the formation of parallel macrosteps works for both initial morphologies, i.e. epitaxial surface and dry etched one, but it is less effective with polished surfaces [6]. This means that surface roughening, either by forming elongated microsteps (epitaxy) or after dry etching, is beneficial to generate the parallel steps. From these observations, we propose a structuration mechanism, schematized in Fig. 7. We assume that the dissolution process is faster at step edges than on the (0001) basal plane due to a higher number of dangling bonds per atom at these step edges (i.e., each atom is bound to the SiC crystal with fewer bonds). This specific dissolution can occur in all the directions (parallel to (0001) plane) both at the undulated and at the flatter areas. As illustrated in Fig. 7a, some Si-C bilayers could be etched from different sides so that their disappearance could be accelerated. This would lead to stable (0001) terraces formation and ultimately to the creation of large terraces by erasing the surface undulation (Fig. 7b, c). Using the same mechanism, the terraces would rapidly extend laterally (i.e. perpendicularly to the [11$\bar{2}$0] direction) either due to the initial elongated undulations of the surface (homoepitaxial layers) or due to a high density of depressions (dry-etched surfaces). This fast lateral propagation of the dissolution front should explain the formation of very long steps which could cross the entire Si-wetted area (up to 12 mm).

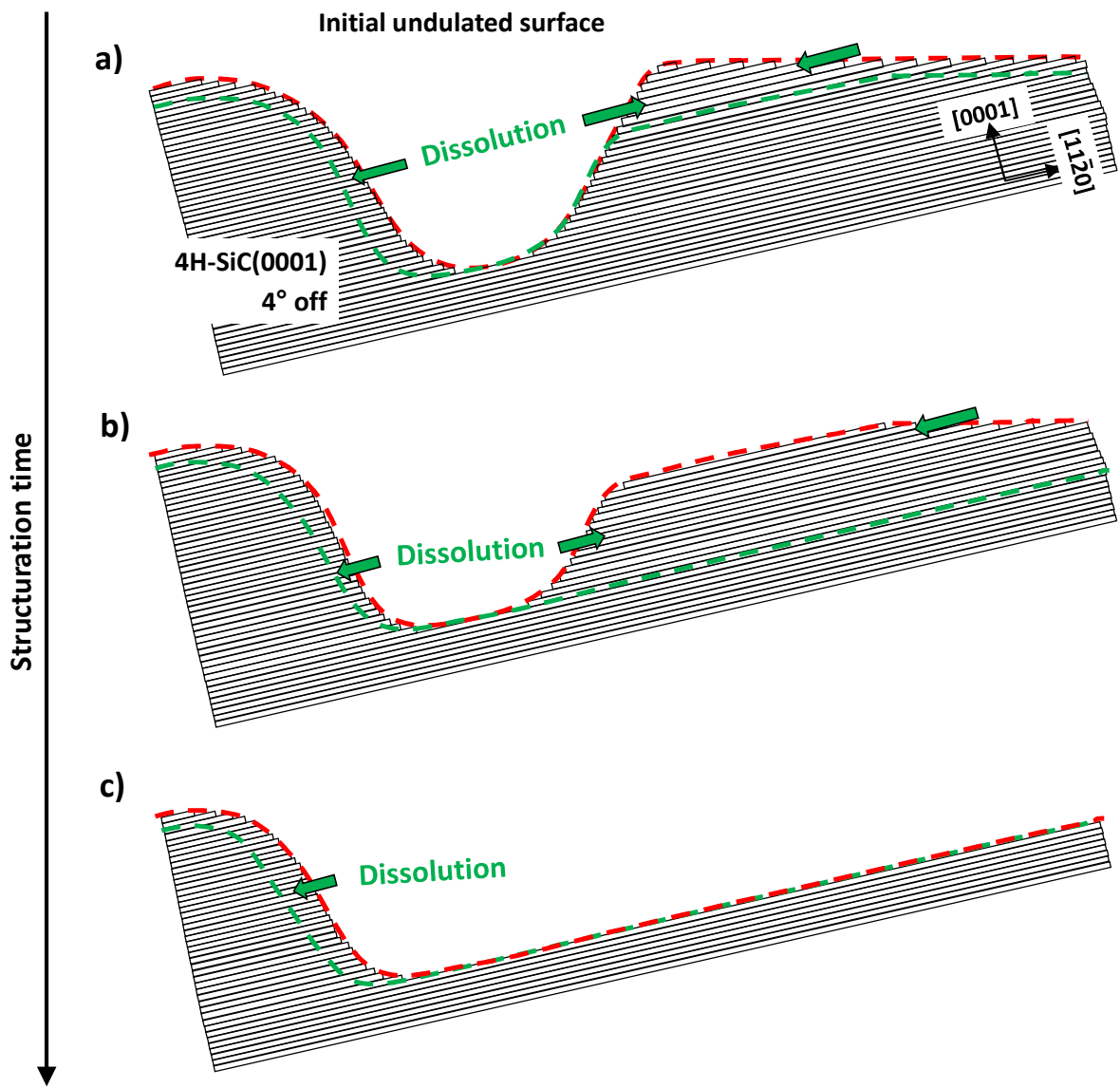

**Figure 7.** Schematic illustration of the proposed mechanism leading to parallel macrosteps formation. Each elongated rectangle represents a Si-C bilayer. In each figure, the red dashed line corresponds to the surface before (further) dissolution; the green dashed line corresponds to the evolution of the surface after further dissolution.

**Summary**

We have investigated the use of patterned surfaces and their effects on 4H-SiC(0001) 4°off surface structuration using a SiC/Si/SiC sandwich approach. Trench patterns should be avoided because of liquid Si spreading difficulties. The use of mesa-patterned samples was more successful and has allowed deeper understanding of the mechanism in play. It should involve preferential etching at Si-C bilayers step edges and fast lateral propagation perpendicular to the $[11\bar{2}0]$ direction. Pattern edges were found to act as obstacles to the lateral propagation of these macrosteps.

**Acknowledgements**

This work has been financially supported by French ANR in the framework of the 19-CE24-0007 "Risemos" project. Authors acknowledge INL (Institute des Nanotechnologies de Lyon) for the use of the mechanical profilometry. The work of M.E. BATHEN was supported by an ETH Zurich Postdoctoral Fellowship.